\documentclass{webofc}

\usepackage[varg]{txfonts}   
\usepackage{hyperref}
\usepackage{url}
\usepackage{float}
\hypersetup{colorlinks=true,citecolor=blue,urlcolor=blue,linkcolor=blue}
\begin{document}
\title{Beam polarization precision requirements for future $e^+e^-$ Higgs factories}
%
%

\author{\firstname{Brendon} \lastname{Madison}\inst{1}\fnsep\thanks{\email{brendon_madison@ku.edu}}
}

\institute{Department of Physics and Astronomy, University of Kansas,
Lawrence, KS 66045, USA
          }

\abstract{We present work on quantifying the minimum requirements for beam polarization precision at future $e^+e^-$ Higgs factories. We find that, under the assumption of a high electron beam polarization ($P_-$) that the positron polarization ($P_+$) is of key importance but for reasons both known and newly discovered. We have discovered that improved positron polarization leads to a less strict requirement on the beam polarization precision for measurements that scale only with the effective polarization, $P_\mathrm{eff}$. Conversely, measurements that scale with the product of beam polarizations, $P_-P_+$, such as those that contain the $eeZ$ or $ee\gamma$ vertex, have their polarization precision demands get more strict as positron polarization increases. We check the polarization precision demands for $10^{-3}$ on the Higgsstrahlung cross-section ($\sigma_{ZH}$) at 250~GeV, $10^{-4}$ on the electron left-right asymmetry ($A_{\rm e}$) at the Z pole, and $10^{-4}$ on the di-photon cross-section ($\sigma_{\gamma \gamma}$) from the Z pole to 3~TeV. We find that, for measurements away from the Z pole, the goals can plausibly be attained if one can achieve precision on beam polarization better than $0.1\%$. For measurements of $A_{\rm e}$ at $10^{-4}$, colliders must do better than $0.02\%$ on beam polarization precision, or determine ways to upgrade their beam polarization values towards unity, where the requirements decrease to better than $0.15\%$.
}
\maketitle
\section{Introduction}
\label{intro}

We will restrict this work to the case of longitudinal polarization; the unaccompanied phrase ``polarization'' will refer to longitudinal polarization. The use of the phrase ``precision'' here refers to the relative uncertainty, $\delta a/a$ for a quantity $a$, while ``uncertainty'' will refer to absolute uncertainty or $\delta a$. In cases where quantities are zero, we will interchange their precision with absolute uncertainties to avoid an undefined value for precision. We also use $P_-$ for the electron beam polarization and $P_+$ for the positron beam polarization. These values will be used as magnitudes in this work, with their signs referred to by $+$ or $-$. These signs can be added to any equation with $P_-$ and $P_+$, but they have not been added in order to keep the notation clear and compact. We have chosen to restrict our investigation to a generic unpolarized collider design, two possible ILC designs with $(P_-,P_+)=(80\%,30\%)$ and $(P_-,P_+)=(80\%,60\%)$, the CLIC/C$^{3}$ designs with $(P_-,P_+)=(80\%,0\%)$, and a ``near Sokolov-Ternov limit'' design for ReLiC, as well as other designs that plan to use the Sokolov-Ternov effect with beam rotators to achieve high values of beam polarization~\cite{LCVision,relic}. We restrict the last case to $(P_-,P_+)=(90\%,90\%)$ as a very optimistic scenario where the Sokolov-Ternov maximum could be reached, and beam rotators and beam transport had very little degradation, thus performing better than the 60-70\% observed at HERA~\cite{Duren:1994wy}.

Work on the design and feasibility of future $e^+e^-$ Higgs factories has shown that longitudinal beam polarization can play an important role in increasing the precision of measurements of couplings, both from the Standard Model (SM) and Beyond the Standard Model (BSM)~\cite{Moortgat_Pick_2008}. Beam polarization also enhances the cross-section of vector boson mediated processes and can change other processes on a case-by-case basis. In many cases, precision may be enhanced over the use of unpolarized beams, simply due to an increase in the statistical precision that the larger polarized cross-sections may yield~\cite{Moortgat_Pick_2008}. Factors like these mean that the precision accessible to a polarized collider of modest instantaneous luminosity, such as $10^{34}\mathrm{cm}^{-2}\mathrm{s}^{-1}$, is comparable to that of unpolarized colliders with much higher instantaneous luminosities, such as $10^{36}\mathrm{cm}^{-2}\mathrm{s}^{-1}$~\cite{deblas2024globalsmeftfitsfuture}. This enhancement is only possible when the beam polarization is measured with a precision sufficient for the demands of a given physical observable. Previous work to estimate these demands on beam polarization precision has set goals of $0.1\%$ for runs related to Higgs boson measurements and $<0.1\%$ for runs related to precision measurements at the Z pole or W pair threshold~\cite{GrahamWW,behnke2007ilcreferencedesignreport,list2021ilcpolarimetry}. Another study found that if one plans to use $e^+e^-\!\to\!\gamma\gamma$ (di-photons) as an integrated luminosity ($\mathcal{L}$) channel, similar levels of beam polarization precision are needed to achieve the integrated luminosity precision goals at various center-of-mass energies~\cite{madison2025integratedluminosity100ppm}. That is to say, if one only uses the opposite-sign beam polarization combinations of $+-$ or $-+$, with no fraction of the same-sign combinations of $++$ or $--$, then a measurement of $\mathcal{L}$ at a precision of $10^{-4}$ will require $\approx10^{-4}$ precision in the measurement of beam polarization.

\subsection{Are Polarimeter and Event Fit Measurements the same?}
For measuring beam polarization, one can use polarimeters, which are instruments that act alongside accelerator structures to reconstruct the beam polarization. These are typically done with Compton polarimeters and have precision in the $10^{-3}$ range~\cite{SBoogert_2009}. The alternative is to use detector information from events that have a well known polarization dependence that can be used to fit the beam polarization. We refer to this approach as the event fit method. Previous studies on event fit methods also find that factors of $10^{-3}$ are feasible~\cite{Marchesini:2011aka}. While these are measuring the same quantity, beam polarization, they are not necessarily the same, especially at the $10^{-3}$ to $10^{-4}$ level of precision that will be discussed here. Two previous studies on beam transport and beam-beam depolarization, respectively, found that sources of bias and uncertainty at the $10^{-4}$ level become relevant at high energy $e^+e^-$ colliders~\cite{Beckmann_2014,Moortgat-Pick:2008zjf}. Therefore, measurements from the interaction point, using event fits, and measurements from the polarimeters may not agree when the beam polarization precision is increased to similar factors of $10^{-4}$. 

The issues at this higher level of precision are not solely for polarimeter measurements; event fitting methods require assumptions of an underlying model that must be evaluated to a similar level of precision~\cite{Blondel:1987wr}. That is to say, if one desires leading order precision on beam polarization, then a similar level of precision is needed in the modeling of the chiral structure of the event type used in the fit. Using leading order calculations from the Standard Model is the typical choice; however, it still poses problems due to the asymmetries that arise in higher order corrections. Studies have shown that electroweak corrections alter the polarization dependence and asymmetry of events used in fitting at the $10^{-3}$ to $10^{-2}$ level, with these corrections generally being larger as the center-of-mass energy increases ~\cite{Berends:1987zz,Bondarenko:2022xmc,Bondarenko:2024txj}. Finding discrepancies in polarimeter measurements, event fit measurements, and the underlying theory and model will be required for any precision physics goals that depend on beam polarization precision. This hunt for discrepancies will also serve as a test of the chiral structure of the Standard Model, beam-beam modeling, and beam transport modeling. 

In the sections that follow, we will derive, with some assumptions to facilitate clearer communication, the minimum requirements for beam polarization precision for various $e^+e^-$ collider designs and measurements.

\section{Beam Polarization Requirements for some Scenarios}
\label{BeamPol}
In this section, we will investigate the minimum beam polarization precision required for various scenarios. We provide table~\ref{tab-meas} as a reference for each measurement, a rough estimate of the beam polarization precision required, and a reference to the table or section with values specific to a given collider design.
\begin{table}[h]
\centering
\begin{tabular}{lll}
\hline
Center-of-mass Energy & Measurement & Req. Beam Pol. Precision \\
\hline
$250~\mathrm{GeV}$ & $\sigma_{\mathrm{ZH}}$ & $\lesssim0.2\%$, table~\ref{tab-ZH} \\
$m_{\mathrm{Z}}$ & $A_{e}$ & $\lesssim 0.02\%$, table~\ref{tab-alr} \\
Any & $\sigma_{\gamma\gamma}$ for $\frac{\delta\mathcal{L}}{\mathcal{L}} \leq10^{-4}$ & $\lesssim0.1\%$, section~\ref{ssec:Lumi} \\
\hline
\end{tabular}
\caption{Collection of precision measurements with strict requirements on the beam polarization precision. We provide rough estimates for the minimum beam polarization precision needed, with more specific values left to the tables referenced.}
\label{tab-meas}
\end{table}

\subsection{General Form for Longitudinal Beam Polarization Error Propagation}
\label{ssec:Gen}
For the sake of communication and reference, we now derive the general form of error propagation for a longitudinal polarization dependent cross-section that has arbitrary coefficients for the beam polarization dependence. A general form for cross-section dependence on asymmetry and polarization terms in $e^+e^-$ interactions can be written
\begin{equation}\label{eq-geneq}
    \sigma = \sigma_0 \left( A_{0} + A_{-}P_- + A_{+}P_+ + A_{+-}P_-P_+ \right)
\end{equation}
in terms of the unpolarized cross-section, $\sigma_0$, and yet-to-be-determined coefficients that represent the unpolarized contribution, $A_{0}$, the electron single-spin contribution, $A_-$, the positron single-spin contribution, $A_+$, and the double-spin contribution, $A_{+-}$. This is equivalent to other formalisms used elsewhere, but it has been rewritten so that we can separate the various polarization dependencies~\cite{Moortgat-Pick:2005jsx}. In fact, one can derive, in the limit of a massless s-channel vector process, that $A_+=A_{LR}=-A_{-}$, the left-right asymmetry used in various references~\cite{Moortgat-Pick:2005jsx}. The coefficients in equation~\ref{eq-geneq} can be derived from theory or from the measured values of the cross-sections. We use existing theoretical calculations of cross-sections, with subscripts to denote the electron and positron helicities, to derive the coefficients
\begin{align}\label{eq-coef1}
    A_{0} = \frac{1}{4\sigma_0}\left(\sigma_{RR} + \sigma_{RL} + \sigma_{LR} + \sigma_{LL} \right) \\
    A_{-} = \frac{\sigma_{RR}+\sigma_{RL}-\sigma_{LR} - \sigma_{LL}}{4\sigma_0}\\
    A_{+} = \frac{\sigma_{RR}-\sigma_{RL}+\sigma_{LR} - \sigma_{LL}}{4\sigma_0}\\ \label{eq-coef4}
    A_{+-} = \frac{\sigma_{RR}-\sigma_{RL}-\sigma_{LR} + \sigma_{LL}}{4\sigma_0}
\end{align}
with empirical relationships from the cross-sections. As a sanity check, we expect $A_0=1$ for a process with no helicity dependence due to how the helicity cross-sections are defined with respect to the unpolarized cross-section. For the sake of error propagation of equation~\ref{eq-geneq}, we derive two helper terms
\begin{align}\label{eq-generr1}
    d_- = \frac{1}{\sigma}\frac{\partial \sigma}{\partial P_-} = \frac{A_{-} + A_{+-}P_+}{A_0+A_-P_-+A_+P_+ +A_{+-}P_-P_+}\\ \label{eq-generr2}
    d_+ = \frac{1}{\sigma}\frac{\partial \sigma}{\partial P_+} = \frac{A_{+} + A_{+-}P_-}{A_0+A_-P_-+A_+P_+ +A_{+-}P_-P_+}
\end{align}
and look for interesting limits and cases. In the unpolarized case, equations~\ref{eq-generr1} and~\ref{eq-generr2} are non-zero if there is a single-spin term. This means that, in the unpolarized $e^+e^-$ collider case, the contributions of beam polarization precision to the cross-section precision may be non-zero.

Lastly, we consider the case in which the beam polarization values are asymmetric, such as in the scenario where there is no positron beam polarization, but the electron beam is highly polarized. In this case, the numerator of equation~\ref{eq-generr1} will depend only on the single-spin electron term, $A_-$. Therefore, in the case of no positron polarization, any process with a zero single-spin contribution will have no dependence on the electron beam polarization uncertainty, $\delta P_-$. Conversely, the numerator for equation~\ref{eq-generr2} will depend on both the single-spin and double-spin terms, as well as the electron beam polarization. The contribution of the positron beam polarization to the precision of the cross-section may be non-zero, even though the positron beam is unpolarized. Therefore, in the case of asymmetric beam polarizations, the form of the single-spin and double-spin coefficients is important. It may result in scenarios where some observables have strict requirements on the beam polarization for the much-less polarized beam or vise versa.

\subsection{Higgsstrahlung Cross-section}
\label{ssec:ZH}
We will now derive the level of precision in beam polarization that is needed for Higgsstrahlung ($e^+e^-\!\to\!ZH$). Since Higgsstrahlung is dominated by the s-channel Z boson contribution, the polarization effects, at leading order, can be derived from this contribution. Starting from equation~\ref{eq-geneq}, the polarization dependent Higgsstrahlung cross-section can be written as
\begin{equation}\label{eq-zh}
    \sigma_\mathrm{ZH} = \sigma_0\left[1 - P_-P_+ + A_-P_- + A_+P_+\right] \approx (1-P_-P_+)\sigma_0(1-P_{\mathrm{eff}}A_\mathrm{\rm e})
\end{equation}
where the double-spin coefficient is $A_{+-}=-1$, the left-right asymmetry for electrons is written as $A_{\rm e}=-A_-=A_+$, and the unpolarized cross-section is written as $\sigma_0$~\cite{Moortgat_Pick_2008}. We use the common notation for the effective polarization
\begin{equation}\label{eq-eff}
    P_\mathrm{eff} = \frac{P_- - P_+}{1-P_-P_+}
\end{equation}
as dependent on the polarization of the electron and positron beams. We note that, for energies near 250~GeV, a common operating point for measuring Higgsstrahlung, the value of $A_{\rm e}$ is $\approx 0.1$, even once NLO QED and weak loop corrections are included~\cite{Arbuzov:2020ghr}. The NLO QED and weak loop corrections to Higgsstrahlung do cause the magnitude of the single-spin coefficients of $A_{-}$ or $A_{+}$ to deviate significantly from $A_{\rm e}$; however, these differences are due to $A_{\rm e}$ increasing in sensitivity to the polar angle as center-of-mass energy increases~\cite{Bondarenko:2018sgg}. For example, at 1000~GeV, a Higgsstrahlung event with a Z emitted at a polar angle of $\pi/2$ has a value of $A_{\rm e}\approx0.08$, while an event with a Z emitted at a polar angle of $\pi/12$ has a value of $A_{\rm e}\approx0.18$. By comparison, two events with the same Z polar angles as before, but now at a center-of-mass energy of 250~GeV, would both see $A_{\rm e}\approx0.1$. Therefore, we will leave it to future studies to determine more specific values of $A_{\rm e}$ to use here and instead focus on the 250~GeV case with $A_{\rm e}\approx0.1$. Using equation~\ref{eq-zh} and the helper terms of equations~\ref{eq-generr1} and~\ref{eq-generr2}, we can derive the helper terms specific to Higgsstrahlung
\begin{align}\label{eq-params1}
    d_- = \frac{-(A_\mathrm{e}+P_+)}{1-P_-P_+ + A_\mathrm{e}(P_+ - P_-)} \\ \label{eq-params2}
    d_+ = \frac{(A_\mathrm{e}-P_-)}{1-P_-P_+ + A_\mathrm{e}(P_+ - P_-)}
\end{align}
which depend on the asymmetry and polarization terms of equation~\ref{eq-zh}. Using equations~\ref{eq-params1} and~\ref{eq-params2}, along with a reference for $A_\mathrm{e}$, we can estimate the minimum level of beam polarization precision needed to reach any precision goal of the Higgsstrahlung process for a polarized electron-positron collider. For the precision goal envisioned for Higgsstrahlung at 250~GeV at various colliders, we assume that one needs at least a contribution from beam polarization of $0.1\%$ or less. An estimate of the minimum precision needed to reach this goal from the precision of the beam polarization of the electron and positron beams can be found in table~\ref{tab-ZH}.
\begin{table}[h]
\centering
\begin{tabular}{lcccc}
\hline
Experiment & Nominal $P_{-}$ & Nominal $P_{+}$ & Req.\ prec. on $P_{-}$ & Req.\ prec. on $P_{+}$ \\
\hline
ILC (80/30)      & $80\%$ & $30\%$ & $0.15\%$ & $0.24\%$ \\
ILC (80/60)      & $80\%$ & $60\%$ & $0.06\%$ & $0.09\%$ \\
CLIC/C$^{3}$ (baseline)  & $80\%$ & $0\%$   & $0.77\%$ & $0.09\%^{*}$ \\
Near ST limit & $90\%$ & $90\%$ & $0.02\%$ & $0.02\%$ \\
Unpolarized & $0\%$ & $0\%$ & $0.67\%^{*}$ & $0.67\%^{*}$ \\
\hline
\end{tabular}
\caption{Minimum required beam polarization precision to keep the \emph{polarization induced} contribution to the Higgsstrahlung cross-section, as measured at 250~GeV, at $0.1\%$ in total. Uncertainties on $P_-$ and $P_+$ are taken uncorrelated and split equally in quadrature.
\\ \footnotesize
$^*$With $P_+=0$ the precision is undefined so the quoted values are uncertainties.
}
\label{tab-ZH}
\end{table}

If we inspect equations~\ref{eq-params1} and~\ref{eq-params2}, we can examine two limits where the precision demands will drastically increase. First, the presence of $1-P_-P_+$ in the denominator will result in colliders with high values of both positron and electron polarization having sharply increasing demands on beam polarization precision. Secondly, the presence of $P_+-P_-$ in the denominator will result in colliders with similar values of beam polarization for both beams also having increasing demands for beam polarization precision. Designs that use the Sokolov-Ternov effect with beam rotators to achieve high values of beam polarization for both beams will result in a worst-case scenario for the requirements on the precision of the measurements of beam polarization. As observed in table~\ref{tab-ZH}, the requirements approach $1\times10^{-4}$. We also expect that, as observed in table~\ref{tab-ZH}, there is a nonzero contribution even in the case of unpolarized beams due to the presence of left-right asymmetry. As such, even unpolarized experiments will need polarimeters or other methods to measure their beam polarization.

\subsection{Left-right Asymmetry at Z Pole}
\label{ssec:ZPole}
For polarization dependent measurements at the Z pole, perhaps the clearest demand for beam polarization precision is from the measurement of the electron left-right asymmetry ($A_\mathrm{LR}$ or $A_{\rm e}$). Previous studies on measurements of $A_\mathrm{e}$ at a polarized run found that various sources of uncertainty make going beyond a precision on $A_\mathrm{e}$ of $10^{-4}$ not feasible~\cite{irles2019complementarityilc250ilcgigaz}. As such, we will choose this as our goal from which to derive the associated minimum precision on beam polarization for the polarized Z pole run. Existing work has shown that
\begin{equation}\label{eq-Alr}
    \frac{\delta A_\mathrm{e}}{A_\mathrm{e}} = \frac{\delta P_\mathrm{eff}}{P_\mathrm{eff}}
\end{equation}
the precision on $A_\mathrm{e}$ is equal to the precision on the effective polarization, originally shown in equation~\ref{eq-eff}. Using the propagation of uncertainty of equation~\ref{eq-Alr}, we derive two helper terms
\begin{align}\label{eq-AlrHelper1}
    a_- = \frac{1}{P_{\rm eff}}\frac{\partial P_{\rm eff}}{\partial P_-} = \frac{1-P_+^2}{(P_- - P_+)(1-P_-P_+)}\\ \label{eq-AlrHelper2}
    a_+ = \frac{1}{P_{\rm eff}}\frac{\partial P_{\rm eff}}{\partial P_+} = -\frac{1-P_-^2}{(P_- - P_+)(1-P_-P_+)}
\end{align}
to compact notation. We use equations~\ref{eq-AlrHelper1} and ~\ref{eq-AlrHelper2} with the original equation~\ref{eq-Alr} to write the dependence of precision on $A_\mathrm{e}$
\begin{equation}\label{eq-AlrPrec}
    \left(\frac{\delta A_\mathrm{e}}{A_\mathrm{e}} \right)^2 = \frac{(1-P_+^2)^2\delta P_-^2 + (1-P_-^2)^2\delta P_+^2}{(1-P_-P_+)^2(P_--P_+)^2}
\end{equation}
from the precision of the beam polarization of each respective beam. Keep in mind that the values of polarization here are signed and, in order to measure left-right asymmetries effectively, one must use opposite signs for $P_-$ and $P_+$. Using equation~\ref{eq-AlrPrec}, we compute the values for the minimum required beam polarization precision, as seen in table~\ref{tab-alr}, under the same previous assumptions that the uncertainty is equally shared between the beams and that they are independent.
\begin{table}[h]
\centering
\begin{tabular}{lcccc}
\hline
Experiment & Nominal $P_{-}$ & Nominal $P_{+}$ & Req.\ prec. on $P_{-}$ & Req.\ prec. on $P_{+}$ \\
\hline
ILC (80/30) & $80\%$ & $30\%$ & 0.01\% & 0.09\% \\
ILC (80/60) & $80\%$ & $60\%$ & 0.03\% & 0.07\% \\
CLIC/C$^3$ (baseline) & $80\%$ & $0\%$ & 0.01\% & 0.02\%$^*$ \\
Near ST limit & $90\%$ & $90\%$ & 0.14\% & 0.14\% \\
Unpolarized & $0\%$ & $0\%$ & 0.01\%$^\dagger$ & 0.01\%$^\dagger$ \\
\hline
\end{tabular}
\caption{Minimum required beam polarization precisions for measuring $A_{\rm e}$ at a precision of $10^{-4}$. We assume uncorrelated beam polarization uncertainties and split the budget equally.  \\ \footnotesize $^*$ When nominal polarization is zero, a relative precision is undefined and we report the required uncertainty instead.\\ \footnotesize $^\dagger$ In the unpolarized beams case one can use forward-backward asymmetry with fermion asymmetry, as in equation~\ref{eq-Afb}, to back-out $A_\mathrm{\rm LR}$ using an existing value of $A_{\rm e}$.}
\label{tab-alr}
\end{table}

For the unpolarized case, the measurement and handling of $A_\mathrm{LR}$ is different from $A_{\rm e}$ because the measurement is based on a calculation dependent on the forward-backward asymmetry ($A_\mathrm{FB}$) and an already measured fermion asymmetry, often $A_{\rm e}$~\cite{swartz1988physics_polarized_electron_beams}. In this case, one typically chooses a final state fermion pair such that
\begin{equation}\label{eq-Afb}
    A_\mathrm{FB}^\mathrm{f} = \frac{3}{4}A_\mathrm{f}\frac{A_\mathrm{LR}-P_\mathrm{eff}}{1-A_\mathrm{LR}P_\mathrm{eff}}
\end{equation}
the forward-backward asymmetry is coupled to the left-right asymmetry via the fermion specific asymmetry ($A_\mathrm{f}$) and the effective polarization. For the sake of notation, we will distinguish the left-right asymmetry that is derived from the forward-backward asymmetry as $A_\mathrm{LR}(A_\mathrm{FB})$. Using equation~\ref{eq-Afb}, we can propagate uncertainty to determine the impact of beam polarization precision. Since we are doing this for the unpolarized case, we can set the polarizations to zero, but only after propagating the uncertainty. In this limit of unpolarized beams, and assuming they are independent, there is a non-zero contribution from the beam polarization precision as
\begin{equation}\label{eq-AfbProp1}
    \frac{\partial A_\mathrm{LR}(A_\mathrm{FB})}{\partial P_-} =     -\frac{\partial A_\mathrm{LR}(A_\mathrm{FB})}{\partial P_+} = (1-A_\mathrm{LR}^2)
\end{equation}
and thus the precision of $A_\mathrm{LR}$ for unpolarized beams using $A_\mathrm{FB}$ and $A_\mathrm{f}$ from the dependence on the beam polarization is
\begin{equation}
    \frac{\delta A_\mathrm{LR}(A_\mathrm{FB})}{A_\mathrm{LR}(A_\mathrm{FB})} = |A_\mathrm{LR}^{-1}-A_\mathrm{LR}|\sqrt{\delta P_-^2 + \delta P_+^2}
\end{equation}
also non-zero. The result of having to extrapolate the left-right asymmetry in this manner is that the precision scales poorer than in the case of polarized beams.

\subsection{Integrated Luminosity using Di-photons}
\label{ssec:Lumi}
For this section, we will focus on the di-photon cross-section. Due to how integrated luminosity is measured from one or multiple well-known processes, one can consider the di-photon cross-section as directly related to the integrated luminosity if one uses di-photons for measuring integrated luminosity. The leading order dependence of the di-photon cross-section on beam polarization
\begin{equation}\label{eq-dipho1}
    \sigma_{\gamma\gamma,\mathrm{LO}} = \sigma_0[1-P_-P_+]
\end{equation}
follows the product of the two beam polarization values. We propagate uncertainty for equation~\ref{eq-dipho1} to obtain the helper terms
\begin{align}\label{eq-dipho2}
    d_- = \frac{1}{\sigma}\frac{\partial \sigma_{\gamma\gamma,\mathrm{LO}}}{\partial P_-} = \frac{P_+}{1-P_-P_+}\\ 
    d_+ = \frac{1}{\sigma}\frac{\partial \sigma_{\gamma\gamma,\mathrm{LO}}}{\partial P_+} = \frac{P_-}{1-P_-P_+} 
\end{align}
and observe that, due to the numerator, in the unpolarized case, the helper terms are zero. Therefore, at leading order, the beam polarization precision does not impact cross-section precision for unpolarized beams.

\begin{figure}[!h]
        \centering
        \includegraphics[width=0.7\linewidth]{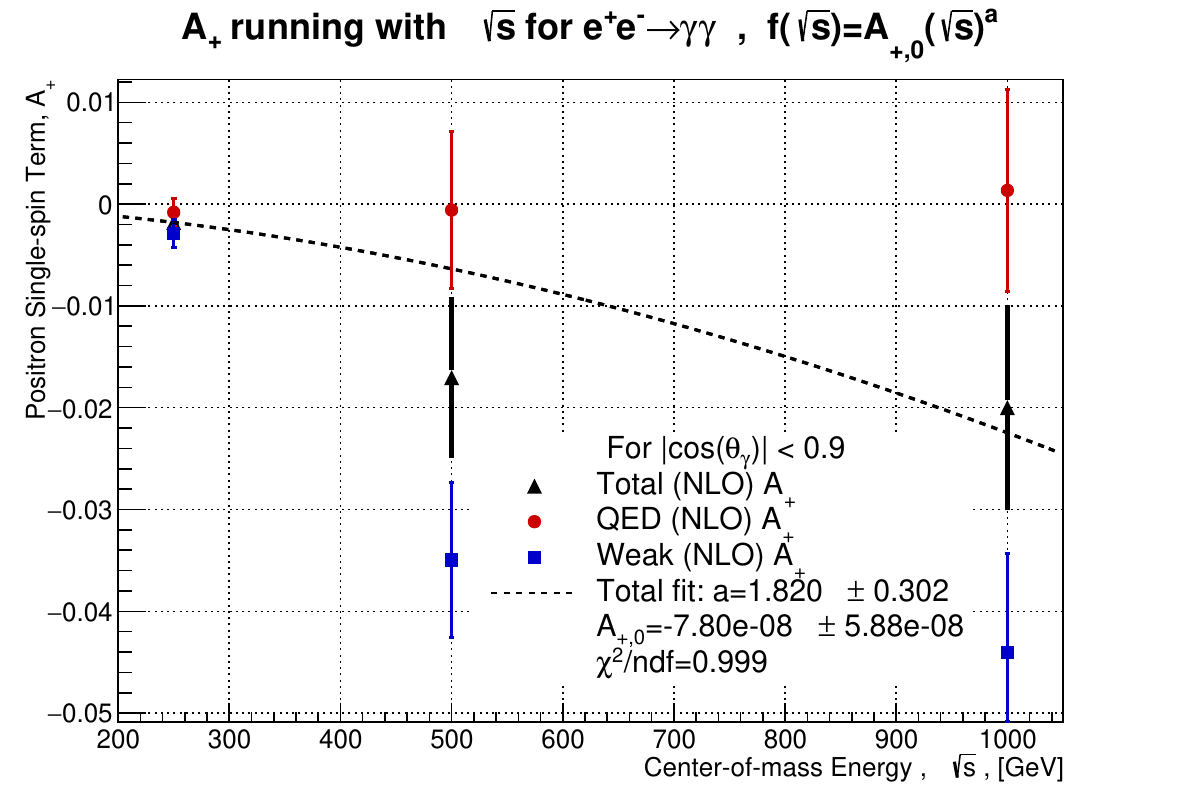}
        \caption{Plot and fit of the single-spin coefficient for the positron beam polarization, $A_+$, as derived in sub-section~\ref{ssec:Gen}. Values used for computation are taken from a reference on one-loop electroweak and QED corrections to $e^+e^-\!\to\!\gamma\gamma$~\cite{Bondarenko:2022xmc}.}
        \label{fig:aplus}
   \end{figure}

\begin{figure}[!h]
        \centering
        \includegraphics[width=0.7\linewidth]{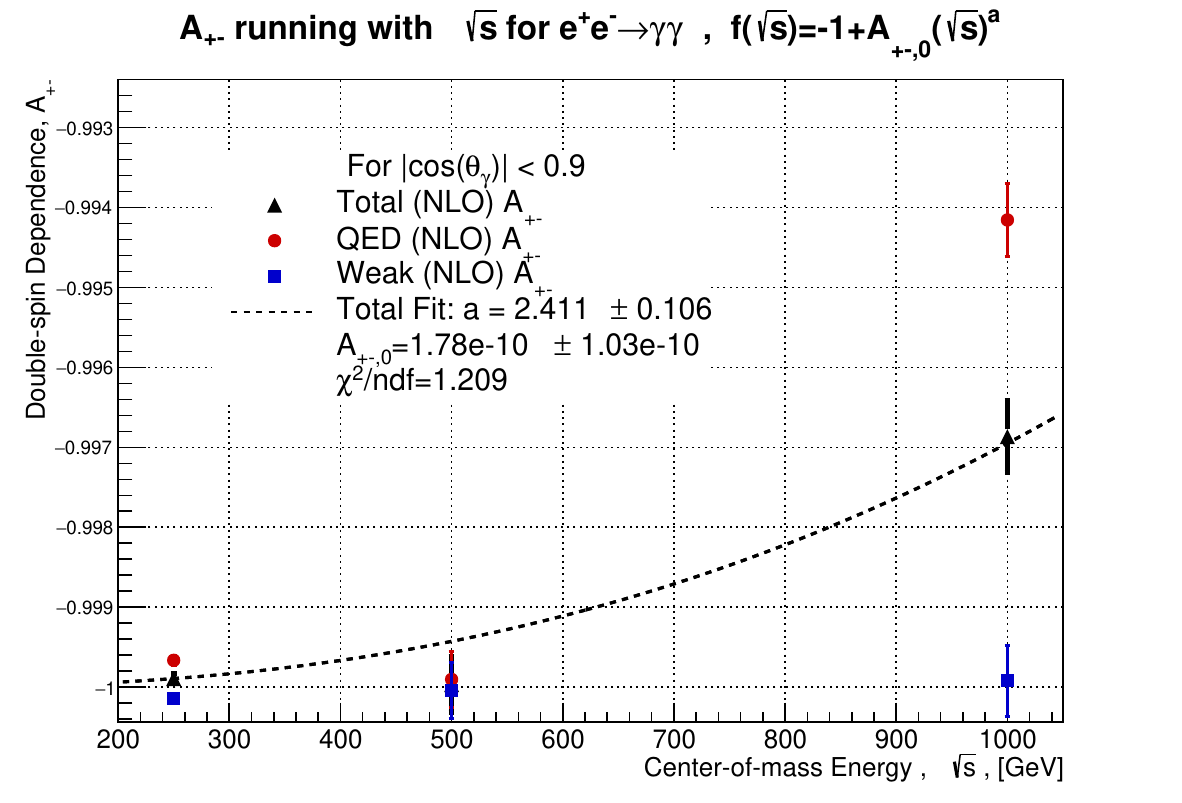}
        \caption{Plot and fit of the double-spin coefficient for electron and positron beam polarizations, $A_{+-}$, as derived in sub-section~\ref{ssec:Gen}. Values used for computation are taken from a reference on one-loop electroweak and QED corrections to $e^+e^-\!\to\!\gamma\gamma$~\cite{Bondarenko:2022xmc}.}
        \label{fig:apm}
   \end{figure}

We now use recent work on one-loop corrections, both electroweak and electrodynamic, to derive the helper terms at next-to-leading-order (NLO)~\cite{Bondarenko:2022xmc}. These theoretical calculations were performed for an angular acceptance of $|\cos(\theta_{\gamma})|<0.9$, which is primarily relevant to wide angle di-photons. As can be seen in the reference, the higher order corrections and their changes to the beam polarization dependence of the di-photon cross-section are much smaller at small angles. We can derive the values of the coefficients in equation~\ref{eq-geneq} using equations~\ref{eq-coef1}--\ref{eq-coef4} and the reference values for center-of-mass energies of 250~GeV, 500~GeV, and 1000~GeV. We also use the theoretical uncertainties quoted and propagate them to the calculations of $A_{+}$ and $A_{+-}$. Examining the cross-sections for the higher order corrections, we find that $A_{-}=-A_{+}$ is within the theoretical uncertainty. This is also expected for reasons discussed in sub-section~\ref{ssec:Gen}. For this reason, we will not plot $A_{-}$ and will assume that $A_{-}=-A_{+}$ for the remainder of this section. We use a one parameter asymptotic fit, with a coefficient that is derived from the reference values, to fit for both $A_{+}$ and $A_{+-}$. We specifically use
\begin{equation}\label{eq-Ap}
A_+ = A_{+,0}\left(\sqrt{s}\right)^{a}
\end{equation}
for $A_{+}$ where $A_{+,0}$ is a constant derived from the data point at 250~GeV and
\begin{equation}\label{eq-Apm}
A_{+-} = -1+A_{+-,0}\left(\sqrt{s}\right)^{a}
\end{equation}
for $A_{+-}$, where $A_{+-,0}$ is also a constant derived from the data point at 250~GeV. We observe that the low energy limit of $A_{+}$ in equation~\ref{eq-Ap} is zero and that the low energy limit of $A_{+-}$ in equation~\ref{eq-Apm} is negative one, which matches the expectation from the leading order case in equation~\ref{eq-dipho1}. As can be seen in figure~\ref{fig:aplus}, the asymptotic fit performs well for modeling the limited data set and suggests that some sort of power-law may effectively approximate the energy running of $A_+$.
In figure~\ref{fig:apm}, we observe that an asymptotic model also performs well for modeling the energy running of $A_{+-}$.
Both figure~\ref{fig:aplus} and figure~\ref{fig:apm} show how the NLO corrections deviate from the leading order values. 

The single-spin deviation is more severe than the double-spin deviation, with fractions of a percent at 250~GeV to a few percent at 1000~GeV. By comparison, the double-spin deviation is roughly one tenth, starting at factors of $10^{-4}$ at 250~GeV and becoming factors of $10^{-3}$ at 1000~GeV. We also observe that the deviation in the single-spin coefficients are dominated by weak contributions, while the deviation in the double-spin coefficient is dominated by QED contributions. This is expected, as QED radiative corrections and loops can cause smearing in the initial state helicities through helicity flips, making the higher order QED corrections less polarization dependent. This is confirmed by examining the values in the reference, which show that the QED corrections have a similar level of correction for all helicity contributions~\cite{Bondarenko:2022xmc}. Conversely, the weak loop contributions can make $\sigma_{LR}-\sigma_{RL}$ and $\sigma_{LL}-\sigma_{RR}$ non-zero due to the preference of the Z boson and W boson for left-handed particles. This causes the single-spin terms to deviate from zero, as they depend on both $\sigma_{LR}-\sigma_{RL}$ and $\sigma_{LL}-\sigma_{RR}$, as derived in equations~\ref{eq-coef1}--~\ref{eq-coef4}.

\begin{figure}[!h]
        \centering
        \includegraphics[width=0.7\linewidth]{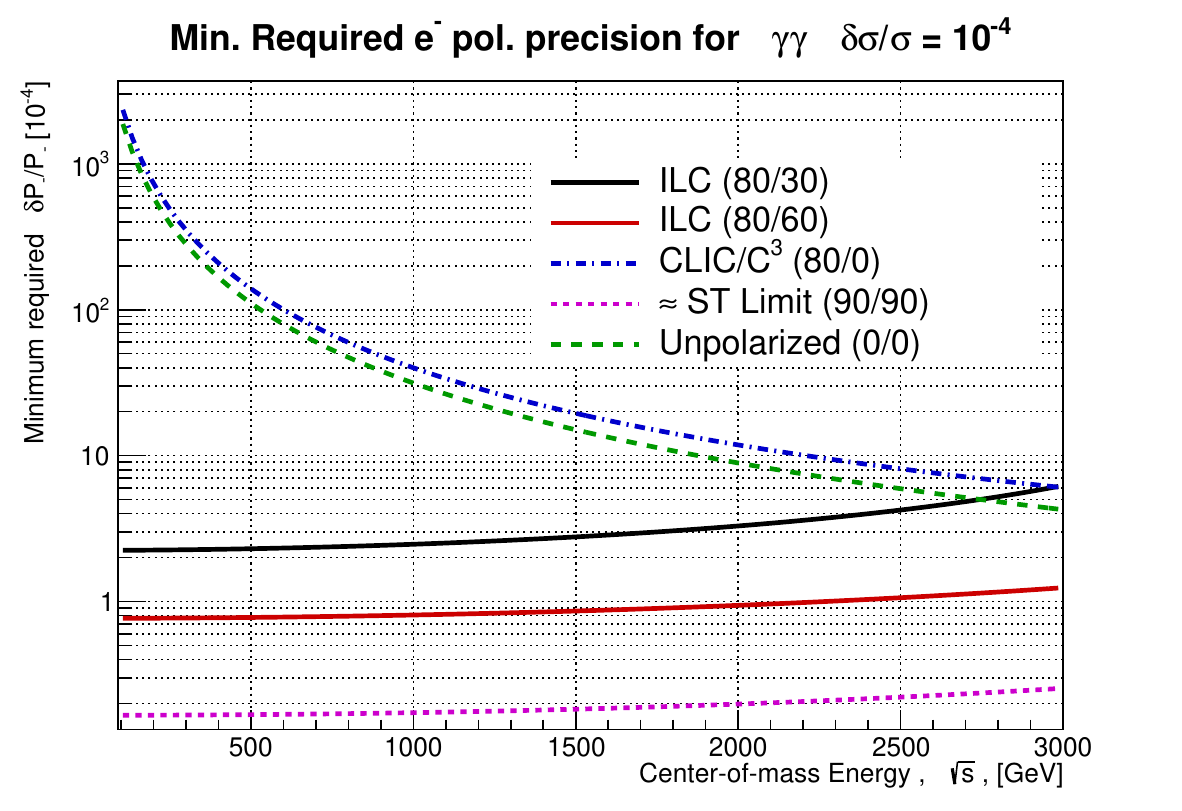}
        \caption{Plot of the minimum electron beam polarization precision required for different collider scenarios for reaching an effect on the di-photon cross-section measurement at a precision of $10^{-4}$. For the colliders with no electron beam polarization the precision quoted is an absolute value, not a relative value, to avoid division by zero.}
        \label{fig:elePrec}
   \end{figure}

\begin{figure}[!h]
        \centering
        \includegraphics[width=0.7\linewidth]{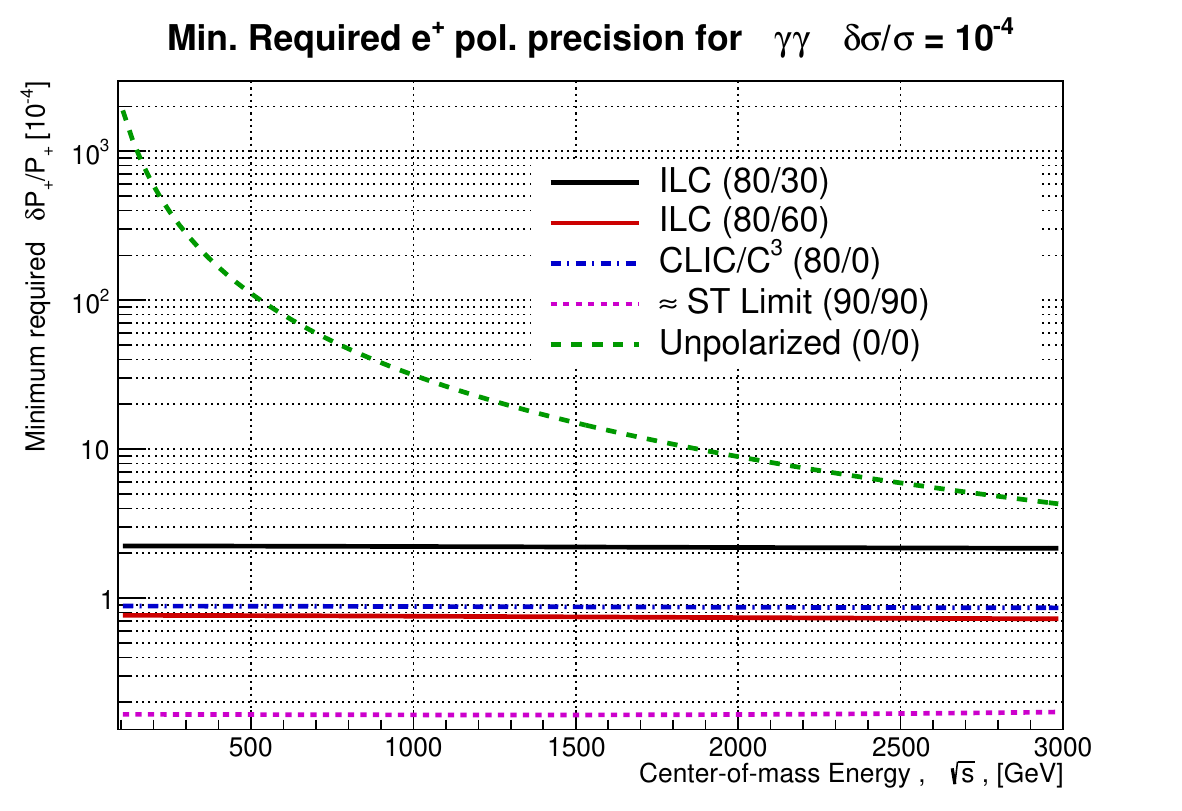}
        \caption{Plot of the minimum positron beam polarization precision required for different collider scenarios for reaching an effect on the di-photon cross-section measurement at a precision of $10^{-4}$. For the colliders with no positron beam polarization the precision quoted is an absolute value, not a relative value, to avoid division by zero.}
        \label{fig:posPrec}
   \end{figure}

\begin{figure}[!h]
        \centering
        \includegraphics[width=0.7\linewidth]{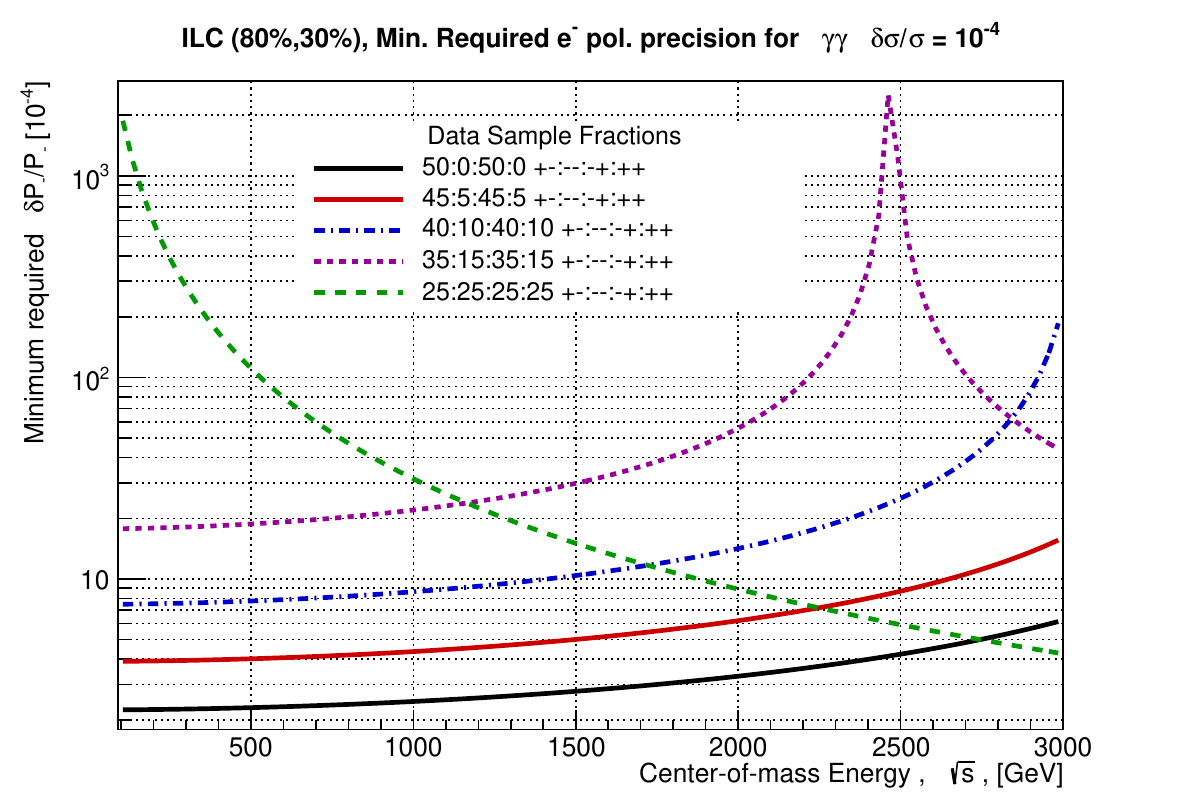}
        \caption{Plot of the minimum electron beam polarization precision required for different amounts of data fractions per polarization sign permutations at the ILC $(P_-,P_+)=(80\%,30\%)$ collider design for reaching an effect on the di-photon cross-section measurement at a precision of $10^{-4}$. For the unpolarized data set, the precision quoted is an absolute value, not a relative value.}
        \label{fig:eleFrac}
   \end{figure}

\begin{figure}[!h]
        \centering
        \includegraphics[width=0.7\linewidth]{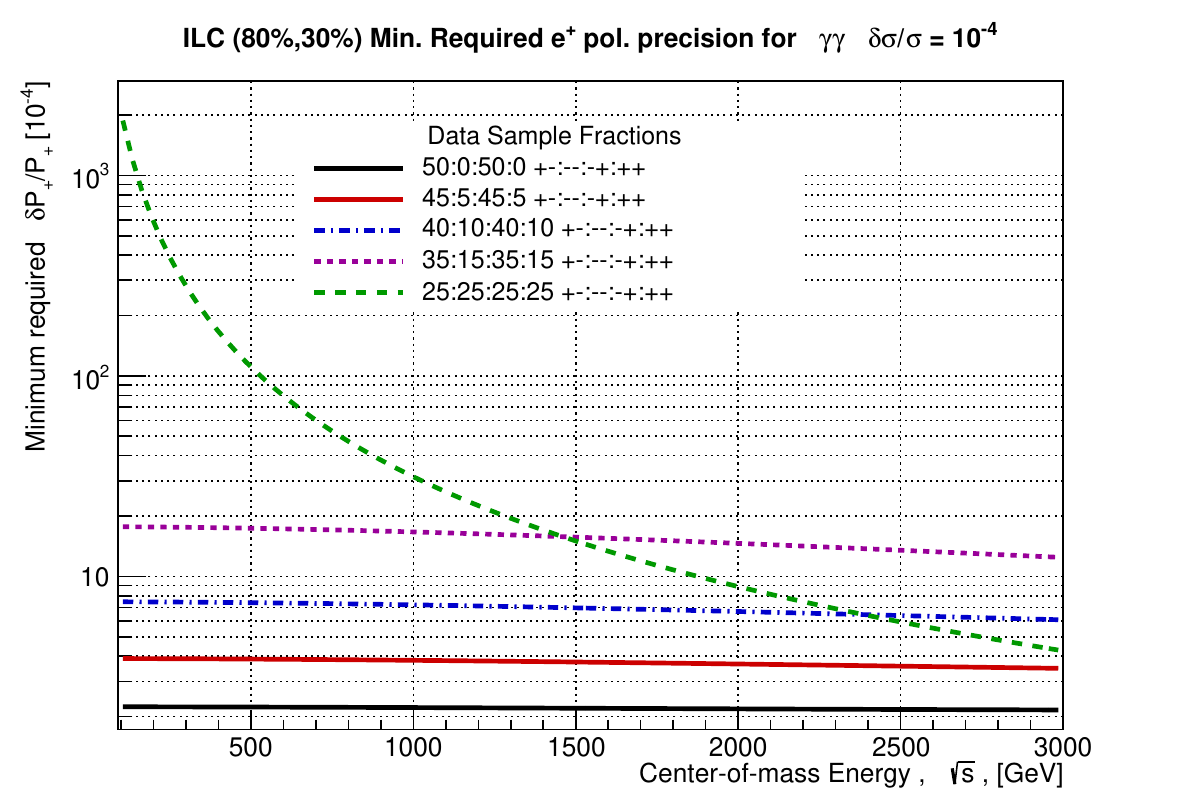}
        \caption{Plot of the minimum positron beam polarization precision required for different amounts of data fractions per polarization sign permutations at the ILC $(P_-,P_+)=(80\%,30\%)$ collider design for reaching an effect on the di-photon cross-section measurement at a precision of $10^{-4}$. For the unpolarized data set, the precision quoted is an absolute value, not a relative value.}
        \label{fig:posFrac}
   \end{figure}

Using the fit results of figures~\ref{fig:aplus} and~\ref{fig:apm}, alongside equations~\ref{eq-generr1} and~\ref{eq-generr2}, we extrapolate the minimum required beam polarization precision at next-to-leading order for each collider beam under the same assumptions used in the previous sections. The results for electron beam polarization precision can be seen in figure~\ref{fig:elePrec}.
The results show how, even for unpolarized beams, the next-to-leading order corrections introduce a non-negligible dependence on the precision of the electron beam polarization. This dependence changes significantly as the center-of-mass energy enters the TeV scale, with slightly polarized beams performing better than unpolarized beams at the highest energies examined here. We find that collider designs with smaller values of positron beam polarization have less strict requirements on the precision of the electron beam polarization. The results for positron beam polarization precision can be seen in figure~\ref{fig:posPrec}.
Similar to electron beam polarization, the unpolarized collider becomes stricter at higher center-of-mass energies. The other designs, which have similarly high levels of electron beam polarization, have roughly constant requirements for the precision of positron beam polarization. For both beams, there is a general trend that higher beam polarization makes the requirements stricter. Therefore, for the di-photon cross-section measurement, it may be more suitable to use moderate values of beam polarization, such as those at the ILC $(P_-,P_+)=(80\%,30\%)$ design. We note that a collider with polarized beams can always recover the unpolarized limit by simply taking equal amounts of data at every polarization permutation. To demonstrate this, we present figures~\ref{fig:eleFrac} and~\ref{fig:posFrac}, which show how different fractions of data at different polarization permutations change the requirements for the precision of beam polarization. 
We observe that even a small fraction of same-sign polarization significantly relaxes the minimum requirements for the precision of beam polarization. We also observe that, in the case of the electron beam, the increase in the precision requirements for the unpolarized case runs counter to the reduction in the precision requirements for the polarized case. Leading to peaks where the center-of-mass energy and data sample fractions minimize the required beam polarization precision. For example, the 70:30 data sample of opposite-sign to same-sign shows that the requirement for beam polarization precision minimizes near 2450~GeV.

In summary, this sub-section presents the first results on extrapolating NLO helicity cross-sections to their associated requirements for the precision of beam polarization at various collider designs. For the goal of a precision of $10^{-4}$ on the di-photon cross-section, we find that the unpolarized collider design is preferable. However, the polarized colliders can make their requirements for the precision of beam polarization easier to achieve by taking small fractions of same-sign beam polarization permutations. Even an 80:20 split for opposite-sign to same-sign data reduces the beam polarization precision requirements to close to $10^{-3}$, which is within the scope of existing polarimeter and event fit methods.

\section{Conclusion}
Within this work, we have shown how the requirements for precision on beam polarization can be derived, as well as what these requirements for various physics goals look like at different $e^+e^-$ Higgs factory designs. We find that, for measurements away from the Z pole, most designs for $e^+e^-$ Higgs factories with polarized beams require better than $0.1\%$ precision in the measurement of the polarization of the beams. For unpolarized colliders, this requirement is relaxed to better than $0.5\%$. At the Z pole, demands for beam polarization precision are strictest from measuring $A_{\rm e}$ to $10^{-4}$, where the requirements are better than $0.02\%$ for polarized colliders. This is with the exception of the case where both beams are highly polarized, where it is lessened to better than $0.15\%$. For unpolarized colliders that wish to back-out left-right asymmetry from forward-backward asymmetry, the scaling is worse and requires beam polarization precision better than $0.01\%$. 

For measurements that become less strict with unpolarized beams, such as the Higgsstrahlung cross-section and di-photon cross-section, these requirements can be relaxed by using larger fractions of same-sign polarized data. For left-right asymmetry measurements, such as $A_{\rm e}$, which are less sensitive at unpolarized colliders, including same-sign data will make the requirements stricter. Therefore, we find two important observations. Firstly, we find that increasing the polarization of the beams is only a straightforward improvement when measuring asymmetries. For cross-section measurements such as Higgsstrahlung or di-photons, increasing beam polarization must be associated with improvements in the measurement of beam polarization. Otherwise, the increased beam polarization may not lead to improved precision in desirable physics goals. Secondly, we find that allowing both the beam polarization and the same-sign data fraction to be tunable between runs allows for the optimization of different measurements. Keeping these values constant results in losing out on potentially significant improvements in precision.

Current methods and technology for measuring beam polarization have placed $0.1\%$ on beam polarization precision as plausible but not concretely attainable. The beam polarization precision requirements for measuring $A_{\rm e}$ to $10^{-4}$ at the Z pole demand $2\times10^{-4}$ at currently attainable values of positron beam polarization. This requirement is the strictest for the scenarios investigated in this work. This can be relaxed to $0.15\%$ if positron and electron beam polarization can be increased towards unity. Otherwise, methods to improve beam polarization measurements to a precision of factors of $10^{-4}$ are needed. This leads to a tension in collider designs, where there must be improvements in positron beam polarization, the measurement of beam polarization, or a combination of the two.

The existing work on increasing positron beam polarization, as well as measuring beam polarization precisely, has been sparse and spread out over the last couple of decades. This, combined with the current uncertainty over whether the existing methods of measuring beam polarization are sufficient, shows a clear need for support for related projects. The preliminary studies and technology development indicate that this may not be a question of possibility, but rather one of supporting a long term study and further technology development. For these reasons, we do not suspect that beam polarization precision will prevent polarized $e^+e^-$ colliders from achieving their goals.

\section{Acknowledgements}

This work is partially supported by the 
US National Science Foundation (NSF) under awards 
NSF~2013007 and NSF~2310030, and it
benefited from the use of the HPC facilities operated 
by the Center for Research Computing at the University of Kansas, which is supported in part by the NSF MRI Award 2117449. We thank Graham Wilson and Jenny List for discussions relevant to this paper.

\global\long\def\bibname{References}%

\bibliographystyle{JHEP}
\bibliography{template}

\end{document}